\documentstyle[prd,aps]{revtex}
\newcommand{\be}{\begin{equation}}
\newcommand{\ee}{\end{equation}}
\newcommand{\bea}{\begin{eqnarray}}
\newcommand{\eea}{\end{eqnarray}}
\def\aprle{\buildrel < \over {_{\sim}}}
\def\aprge{\buildrel > \over {_{\sim}}}
\begin{document}
\draft
\input epsf
\twocolumn[\hsize\textwidth\columnwidth\hsize\csname
@twocolumnfalse\endcsname 

\title{Baryogenesis via neutrino oscillations}
\author{E. Kh. Akhmedov${}^{(a,b)}$ V. A. Rubakov${}^{(c,a,d)}$ 
and A. Yu. Smirnov${}^{(a,c)}$}
\address{${}^{(a)}${\em The Abdus Salam International Centre for
Theoretical Physics, 
I-34100 Trieste, Italy}}
\address{${}^{(b)}${\em National Research Centre Kurchatov Institute,
Moscow 123182, Russia}}
\address{${}^{(c)}${\em Institute for Nuclear Research of the Russian
Academy of Sciences, Moscow 117312, 
Russia}}
\address{${}^{(d)}${\em Institute for Cosmic Ray Research, University
of Tokyo, Tanashi, Tokyo 188, Japan}}

\date{March 5, 1998}
\maketitle
\begin{abstract}
We propose a new mechanism of leptogenesis in which the asymmetries 
in lepton numbers are produced through the CP-violating oscillations of
``sterile'' (electroweak singlet) neutrinos. The asymmetry is communicated 
from singlet neutrinos to ordinary leptons through their Yukawa couplings.
The lepton asymmetry is then reprocessed into baryon asymmetry by electroweak 
sphalerons. We show that the observed value of baryon asymmetry can be 
generated in this way, and the masses of ordinary neutrinos induced by the
seesaw mechanism are in the astrophysically and cosmologically interesting 
range. Except for  singlet neutrinos, no physics beyond the Standard Model
is required.
\end{abstract} 
\pacs{PACS: 98.80.Cq, 14.60.St    \hskip 2 cm IC/98/22, ~INR-98-14T 
\hskip 2cm hep-ph/9803255}
\vskip2pc]

1. The origin of the excess of baryons over anti-baryons
in  the Universe remains one of the fascinating problems
of particle physics and cosmology. 
A number of mechanisms have been proposed to date to explain 
this asymmetry (for recent reviews see, {\it e.g.}, \cite{RS}). 
One of the simplest possibilities, suggested by Fukugita and Yanagida
\cite{FY}, is that the baryon asymmetry has originated from 
physics in the leptonic sector. Namely, it was assumed that
at temperatures well above the electroweak scale, 
lepton asymmetry  was produced, which was then  
reprocessed into the baryon asymmetry by non-perturbative electroweak 
effects \cite{KRS} -- sphalerons \cite{KM}.
According to ref. \cite{FY} the lepton asymmetry  is generated
in out-of-equilibrium, CP- and lepton number non-conserving decays of
heavy Majorana neutrinos (for recent discussions see, {\it e.g.},
ref. \cite{Buch} and references therein).

In this Letter we propose 
a new  realization  of baryogenesis through leptogenesis which
also makes use of the electroweak reprocessing of the lepton number
into the baryon number. Like the Fukugita--Yanagida mechanism, our
proposal requires only mild extension of the Standard Model by
introducing ``sterile'' ({\it i.e.}, electroweak singlet) heavy
neutrinos. However, our mechanism of leptogenesis  is entirely 
different from that of ref. \cite{FY}: 
we suggest that asymmetries in lepton numbers were generated due to 
oscillations of these singlet neutrinos  and their interactions with 
ordinary matter in the early Universe.  Moreover,  
the novel feature of our 
scenario is that the total lepton number is not violated in 
these oscillations and/or interactions; an important ingredient is
separation (rather than generation) of lepton  number, {\it i.e.}, its
redistribution between different species of singlet neutrinos.  

For this reason  we do not necessarily require that singlet 
neutrinos be Majorana particles; Dirac ``sterile'' neutrinos are equally  
suitable  (and even  better in some respect) for our mechanism to work. 
Furthermore, in our case the values of the masses and couplings 
of singlet neutrinos are very different from those 
of ref. \cite{FY}. \\


2.  Let us consider the Standard Model extended by adding 
three types of Majorana neutrinos $N_a$, $a=A,B,C$ which interact with  
other particles only through their Yukawa couplings \cite{*}. 
The corresponding Lagrangian can be  written in the ``Yukawa basis'' 
(where the matrix of Yukawa coupling
constants has been diagonalized) as follows, 
\[
{\cal L}=
\bar{N}_{Ra} i \partial\!\!\!/N_{Ra} + h_{a}\,
\bar{l}_{a} N_{Ra}\Phi 
+
\frac{M_{ab}}{2}\,N_{Ra}^{T} C N_{Rb}
+h.c. ~.  
\]  
Here $N_{Ra}$ are  right-handed components of $N_a$, $h_{a}$
are  (real) Yukawa coupling constants,
$l_a$ are three independent linear combinations of 
usual leptonic doublets, $\Phi$ is the Higgs doublet,
and $M_{ab}$ is the mass matrix.

We suggest that  the baryogenesis proceeds in the following way: 

(i) In the course of the evolution of the Universe,  singlet neutrinos
are produced through their Yukawa couplings. 
The production mechanism of singlet neutrinos conserves 
CP, {\it i.e.} for each type 
equal numbers of particles and antiparticles 
(particles of opposite helicities in the Majorana case) are produced. 

(ii) Once created, singlet neutrinos  oscillate, and 
also interact with ordinary matter. None of these processes violates the
total lepton number $L^{tot}=L + L_A + L_B + L_C$, where $L$ is the 
usual lepton number (we assume that Majorana masses are small
enough, see below). However, CP is not conserved due to mixing in the
singlet neutrino sector. 
Therefore the initially created state with individual lepton 
numbers $L_A=L_B=L_C=0$ evolves through the oscillations 
into a state in which $L_A\ne 0$, $L_B\ne 0$, $L_C\ne 0$  
but still $L^{tot}=0$. That is, the total lepton number gets unevenly 
distributed between different species.

(iii) Singlet neutrinos communicate their lepton asymmetry to  ordinary 
neutrinos and charged leptons through their Yukawa couplings. We assume that 
the Yukawa couplings of $N_A$, $N_B$ and $N_C$ have certain hierarchy,
so that neutrinos of at least one type, $N_A$, come into thermal 
equilibrium before the  time $t_{EW}$ at which sphalerons become 
inoperative (the corresponding  temperature is $T_{EW}\sim 100$ GeV) 
and those of at least one other type, $N_C$, do not equilibrate by
$t_{EW}$. The neutrinos of the third type, $N_B$, may or may
not come into thermal equilibrium by $t_{EW}$. To be specific, we discuss 
mostly the case when the Yukawa coupling of $N_B$ is relatively large 
so that $N_B$ equilibrate at temperatures exceeding $T_{EW}$ (the opposite 
case is treated in a similar way). 
In this case the lepton numbers $L_A$ and $L_B$ are communicated to
the ordinary leptons 
before $t_{EW}$ whereas $L_C$ is not. Therefore
(a fraction of) $L_A+L_B$ is reprocessed into baryon asymmetry by 
electroweak sphalerons; the lepton number $L_C$ is transfered to the 
active leptons only after the sphalerons have already switched off,
so it has no effect on the baryogenesis. 

Let us stress that if not for the sphaleron freezing effect, 
no  baryon asymmetry would have been obtained: all three singlet
neutrino species eventually transform into ordinary leptons, and since
$L^{tot}=0$, no net lepton charge would have been
generated in the sector of ordinary leptons. Hence, the requirement that
$N_C$ do not get into thermal equilibrium before $t_{EW}$ is crucial for
our mechanism.\\ 


3. The system of three types of singlet neutrino of a given momentum
$k(t) \propto T(t)$, interacting with cosmic plasma, is described by
$3\times 3$ density matrix $\rho(t)$ which obeys the evolution equation
\cite{SR}
\begin{equation}
  i\frac{d\rho}{dt} = [\hat{H},\rho]
  -\frac{i}{2} \{ \Gamma, \rho \}
   +\frac{i}{2} \{ \Gamma^{p}, 1 - \rho \}
\label{rho}  
\end{equation}
Here $\hat{H}$ is Hermitean effective Hamiltonian, $\Gamma$
and $\Gamma^{p}$ are destruction and production rates, respectively,  
and we have neglected the relaxation 
due to elastic scattering, see below.  
In what follows we will use the approximation of
Boltzmann statistics for order-of-magnitude analysis; in this approximation
the last term in eq.(\ref{rho}) becomes merely $i\Gamma^{p}$.

In the Yukawa basis one has at $T \gg M_{a}$ 
\begin{equation}
     \hat{H} = V(t) 
      + U\frac{\hat{M}^2}{2 k(t)}U^{\dagger}\,.
\label{H1}  
\end{equation}
Here $U$ is the mixing matrix  which relates the Yukawa states and the 
mass eigenstates $N_i$ ($i = 1, 2, 3$): $N_a = U_{ai} N_i$
(we assume that mixing is small, and choose the standard 
parametrization for $U$ \cite{PDG});
$\hat{M}^2=\mbox{diag}~(M_1^2,M_2^2,M_3^2)$ is 
the matrix of mass eigenvalues; $k(t)$ is the neutrino   
momentum which depends on time due to
the expansion of the Universe. The CP violation 
in the system is described by the CP-odd phase $\delta$ in the mixing 
matrix $U$. 
   
The medium effects 
are described by real potential $V$ and rates 
$\Gamma$ and $\Gamma^{p} = \exp(-k/T) \Gamma$ 
\cite{Weld2} whose matrices are diagonal
in the Yukawa basis, 
\[   V = \mbox{diag}~(V_A, V_B, V_C)\,,\;
 \Gamma = \mbox{diag}~(\Gamma_A,\Gamma_B, \Gamma_C)\,
\]
The potentials $V_a$ are due to the coherent
forward scattering processes,
the main contribution coming from
the 1-loop self energy diagrams with ordinary lepton and Higgs doublets in
the intermediate states. For our estimates we use their thermal averages
which at temperatures above 
$T_{EW}$ are \cite{Weldon}
\be
V_a=\frac{1}{8}\,h_a^2 \,T\,.
\label{V2}
\ee
Consider now the rates $\Gamma_a$.
We will be interested in the temperatures far exceeding the masses of
singlet neutrinos. Therefore the rates of the $1\leftrightarrow 2$ 
reactions which correspond to the absorptive parts of the self energy  
diagrams are suppressed by the factor $M_a/T$ and 
$2\leftrightarrow 2$ reactions are more important. 
The main contributions to $\Gamma_a$ come from the 
Higgs exchange reactions $Q_L N_{Ra}\leftrightarrow t_R\,l_a$, 
$t_R^{~c} N_{Ra}\leftrightarrow Q_L^{~c}\,l_a$ and 
$l_a^{~c} N_{Ra}\leftrightarrow t_R\,Q_L^{~c}$ where $Q_L$ is the
third-generation quark doublet.
The result \cite{Luty} for the average destruction rates
at $T\gg M_a$ (corrected to include the color 
factor for quarks) is 
\be
\Gamma_a\simeq \frac{9 h_t^2}{64\pi^3}\, h_a^2 \, T\, . 
\label{Gammas}
\ee
Here $h_t\simeq 1$ is the top quark Yukawa coupling. The rates 
of the elastic $2\to 2$ scattering processes are proportional to 
$h_a^2 h_b^2$ instead of $h_a^2 h_t^2$. We will need very small
Yukawa couplings of singlet neutrinos, so the 
elastic processes can be safely neglected. 

At late times the lepton asymmetry is stored in the 
least interacting species of singlet neutrino, which at 
$t\sim t_{EW}$ coincides with the 
mass eigenstate $N_3 \approx N_C$. 
The conditions that $N_1 (\approx N_A)$ and $N_2 (\approx N_B)$ come into
thermal equilibrium before the time $t_{EW}$, while $N_3$ do not, are 
\be
\Gamma_{1,2}(T_{EW})> H(T_{EW})\,,\quad 
\Gamma_{3}(T_{EW})< H(T_{EW})\,,
\label{cond1}
\ee
where $H(T)=T^2/M_{\rm Pl}^*$ is the Hubble parameter, $M_{\rm Pl}^*
\equiv M_{\rm Pl}/1.66 \sqrt{g_*}\simeq 10^{18}$ GeV, 
and $\Gamma_{3} \approx 
\Gamma_C + s_{13}^2 \Gamma_A + s_{23}^2 \Gamma_B$. Here  
$s_{13}\equiv\sin\theta_{13}$ and $s_{23} \equiv \sin \theta_{23}$ 
determine the admixtures of $N_A$ and $N_B$ in $N_3$.  

The conditions (\ref{cond1}) translate into bounds on the Yukawa 
couplings, 
\be
h_{A,B}^2 > 2\cdot 10^{-14}\,,\quad h_{C}^2 < 2\cdot 10^{-14}\,,
\label{cond2}
\ee
and on mixing angles,
\be
    s_{13}^2 < 2\cdot10^{-14}  h_A^{-2}\, ,  \;\;
 s_{23}^2 < 2\cdot10^{-14} h_B^{-2}~. 
\label{sines}
\ee
Eq. (\ref{cond2}) implies a certain hierarchy between the Yukawa 
couplings, which, however, need not be very strong.\\


4. As described above, to find the baryon asymmetry we should calculate
the  asymmetry $\Delta_L (t_{EW})$ which was communicated to usual leptons 
by the time $t_{EW}$ at which sphalerons switch off. This asymmetry
emerges because singlet neutrinos $N_3$ do not transfer their asymmetry to 
active neutrinos by the time $t_{EW}$ due to smallness of $h_C$, $s_{13}$  
and $s_{23}$. Since the total lepton number is conserved in all processes of 
interest, we have $\Delta_L (t_{EW}) = - \Delta_3 (t_{EW})$ 
(up to a factor of order one that accounts for the 
distribution of the asymmetry between $B$, $L$ and $L_A + L_B$; we will 
not write this factor in formulas below),  where $\Delta_3 (t_{EW})$
is the asymmetry stored in $N_3$.
The asymmetry $\Delta_3 (t_{EW})$ can be found as follows. Let $S(t,t_0)$
be the evolution matrix  corresponding to the operator
$\tilde{H} = \hat{H} - (i/2) \Gamma$ 
(notice that $S(t,t_0)$ is not unitary since this operator 
is non-Hermitean). 
The density matrix $\rho(t)$ can be expressed through $S(t,t_0)$ assuming
$\rho(t_i) = 0$ where $t_i$ is the time at which the production of singlet 
neutrino begins \cite{fn2}. 
The ratio of the number density of 
$N_3$, $n_3 = \rho_{33}$, 
to the equilibrium density of one spin degree of freedom 
at time $t_{EW}$ is 
\be
\frac{n_3(t_{EW})}{n_{eq} (t_{EW})} = 
 \sum_{a, b} 
\int_{t_i}^{t_{EW}} dt_0~ \Gamma_a (t_0)
|U^{\dagger}_{3 b} S_{b a}(t_{EW},t_0)|^2 ~,
\label{P}
\ee
The asymmetry $\Delta_3 (t_{EW})$ is the
CP-odd part of the quantity (\ref{P}).

The integration over the
production time $t_0$ can be performed in a closed form. Indeed,
the matrices $S$ and $S^\dagger$ obey 
$\partial_{t_0}S = i S \tilde{H}(t_0)$, 
$\partial_{t_0}S^{\dagger} = 
-i \tilde{H}^{\dagger}(t_0)S^{\dagger}$. From these equations 
and $\Gamma  =  i (\tilde{H} - \tilde{H}^{\dagger})$, 
one finds  that $\partial_{t_0}(S S^\dagger) 
= S\Gamma (t_0) S^{\dagger}$.  
Using this relation one can readily perform 
the integration over the production time $t_0$ in 
Eq. (\ref{P}),
\[
\frac{n_3 (t_{EW})}{n_{eq} (t_{EW})} =
    1 - \left[S^M (t_{EW}, t_i) S^M(t_{EW},
t_i)^{\dagger}\right]_{33}~. 
\] 
Here $S^M \equiv U^{\dagger} S U$ is the evolution matrix in the 
mass eigenstate basis. The CP-odd part of this expression 
determines the asymmetry transfered to usual
leptons by $t_{EW}$ (and hence the generated baryon asymmetry)
\bea
\Delta_L(t_{EW}) \equiv (n_L - n_{\bar{L}})/n_{\gamma}= -
\Delta_3 (t_{EW}) \nonumber \\
  = {1 \over 2} \sum_i |S^M_{3i}(t_{EW},t_i)|^2_{CP-odd}~, 
\label{DeltaL1}
\eea
where the factor $1/2$ accounts for two helicity states of photon.
The production of singlet neutrinos starts at
very early times, 
so we set $t_i=0$. Since $T_{EW}$ is much smaller than
all relevant energy parameters in the problem,
one can formally let $t_{EW}\to \infty$ in actual calculations.\\


5. The lepton asymmetry is  produced mainly at the epoch $t_L$ when the 
differences of the eigenvalues of the Hamiltonian $\Omega_{ij} \equiv 
\Omega_i - \Omega_j$ become of order of the Hubble parameter $H$: 
$\Omega_{ij}(t_L)  \sim 1/t_L$. Indeed, at  $t \ll t_L$ the 
 elements of $\rho(t)$ essentially stay constant, whereas at 
$t \gg t_L$ they undergo fast oscillations and due to averaging effects 
the  asymmetry is strongly suppressed. 

In what follows we will present the results for the most interesting
range of the parameter space  
where the mass differences are relatively large,   
$\Delta M^2 \gg$  $(h_{a b}^2/8)^3 M^{* ~2}_{Pl}$  
($h_{a b}^2\equiv h^2_a - h^2_b$), 
$\Delta M^2$ being the typical value of $\Delta M_{ij}^2$. 
In this case  the  mass terms dominate 
over potentials in $\hat{H}$ at the
epoch $t_L$,  so that  $\Omega_{ij} \sim \Delta M_{ij}^2 /2 T$ and    
the leptogenesis temperature is
$T_L  \equiv T(t_L) \sim  (M^{*}_{Pl} \Delta M_{ij}^2)^{1/3}$.   
At $t = t_L$ we have
\be 
\frac{|V_{ab}|}{H} = 
\frac{h_{ab}^2}{8}
     \left(\frac{M^{*\,2}_{Pl}}{\Delta M^2}\right)^{1/3}
\equiv \lambda   \ll 1.
\label{n1*}
\ee
This means that the potentials $V_a$ and rates $\Gamma_a$
can be treated perturbatively  
with $\lambda$ being the expansion parameter.
The lepton asymmetry (\ref{DeltaL1})
appears in the third order of perturbation
theory, and therefore 
is suppressed by the cube of $\lambda$. 
This can be seen in the mass eigenstate basis, 
where the potential has the form  $U^{\dagger} V U$. Indeed, 
$\Delta_L$  is a CP - violating observable, so it should 
be proportional to the invariant 
$J=s_{12}\,c_{12}\, s_{13}\, c_{13}^2\, s_{23}\, c_{23}\, \sin\delta$;
this invariant can be collected 
from $[U^{\dagger} V U]^3$ only. 
It is clear from eqs. (\ref{P}) and (\ref{DeltaL1}) that $\Delta_L$ 
vanishes  in the limit $\Gamma \to 0$, so one expects it to be
proportional also  to
$\sin \phi \equiv \Gamma_a /2 V_a \simeq  2\cdot 10^{-2}$. Therefore, up
to a numerical constant  we have an estimate 
\be
\Delta_L \sim  J \lambda^3 \sin \phi \,.
\label{estimate}
\ee

The calculations are simplified if 
$|\Delta M_{13}^2|\ll$  $|\Delta M_{12}^2|,|\Delta M_{23}^2|$. 
In this case, using Eq. (\ref{DeltaL1}), we obtain
\be 
\Delta_L \simeq  
\frac{\left[\Gamma \left(\frac{1}{3}\right)\right]^3}{384} 
J \sin \phi 
\frac{h_{AC}^2 h_{AB}^2 h_{BC}^2 \cdot M^{*\, 2}_{Pl}}
{(\Delta M^2_{13})^{1/3} (\Delta M^2_{12})^{2/3}}~, 
\label{n3*} 
\ee
where $\Gamma(1/3) \simeq 2.68$ .
The estimate (\ref{estimate})
as well as the formula (\ref{n3*}) are valid both in the case when $N_2$ 
equilibrate before $t_{EW}$  and in the opposite case. 

The asymmetry increases when the 
parameter $\lambda$ approaches 1, {\it i.e.}, the maximal effect for given  
$h_a$ is expected when 
$\Delta M^2 \sim (h_{ab}^2/8)^3  M^{*\, 2}_{Pl}$. 
Notice, however, that these and smaller values of $\Delta M^2$ 
correspond to singlet neutrinos strongly degenerate in mass. \\


6. Let us present constraints on the parameters of singlet 
neutrinos and discuss the value of the asymmetry. 

(i) To be consistent with the standard mechanism of nucleosynthesis, 
all singlet neutrinos, including the most weakly interacting one $N_3$,
should decay before the nucleosynthesis epoch. The decay of 
$N_3$ at $T\ll T_{EW}$ occurs due to its mixing with ordinary
neutrino. 
Requiring that the decay rate of $N_3$ exceeds the inverse  
lifetime of the Universe at temperatures of order of a few MeV,
and recalling Eq. (\ref{cond2}), we obtain a lower bound on the mass, 
$M_3\aprge 1$ GeV. Alternatively, for $M_3\gg 1$ GeV, we have a lower
bound on the Yukawa constant,
$h_3^2 \aprge 10^{-16} (1~{\rm GeV}/M_3)^3 $ where 
$h_3^2 \simeq h_C^2 + s_{13}^2 h_A^2 + s_{23}^2 h_B^2$.

(ii) If the mass of $N_3$ is close to 1 GeV, and/or
the Yukawa couplings of $N_A$ and $N_B$ are close to the bound
(\ref{cond2}), the decays of singlet neutrinos may
lead to the reheating of the Universe after the electroweak 
epoch (but before nucleosynthesis), and hence to the dilution of the
baryon asymmetry. This reheating is rather modest, however: given the
constraints already imposed, the entropy density may increase at most
by a factor of 10.  In this case the baryon asymmetry 
produced before the reheating should be an order of magnitude larger 
than the observed one. 

(iii)  Baryon and lepton asymmetries should not be 
washed out before $T=T_{EW}$ by  the
Majorana mass itself. At $T \gg M_A $ the lepton number equilibration rate 
is suppressed with respect to
the lepton charge conserving rate $\Gamma_A$,
given by Eq. (\ref{Gammas}), by a factor
$M_A^2/T^2$,
so we have to require that $\Gamma_A(T_{EW})(M_A^2/T_{EW}^2) \ll
H(T_{EW})$, and similarly for $N_B$.
The parameters of Majorana singlet neutrinos should therefore satisfy 
\be
1\; {\rm GeV} \aprle M_a \ll 100\; {\rm GeV}\,,\:\;\;
h_A^2\, , \, h_B^2 \ll 10^{-10}\,. 
\label{masses}
\ee 
The upper bounds here do not apply to Dirac singlet neutrinos.

The Lagrangian of the model leads, via the see-saw mechanism,
to the generation of masses of the light (active) neutrinos:
$m_{\nu_a} \equiv m_{a} =  h_a^2 v^2/M_a$, where $v$ is  the Higgs vacuum 
expectation value and $\nu_a$ are 
mass eigenstates -- combinations 
of $\nu_e$, $\nu_{\mu}$ and $\nu_{\tau}$. 
The constraints (\ref{cond2}) and (\ref{masses}) imply that
the mass of the heaviest active  neutrino is in the range
$m_{a} = (10^{-2}  - 10^3)$ eV. From  the cosmological bound 
$m_{a} \aprle 10$ eV  
we get {a constraint which is} somewhat stronger than Eq. (\ref{masses}),
$h_A^2 \aprle 10^{-11}$.  
For  the lightest active neutrino, the constraints 
(\ref{cond2}) and (\ref{masses}) 
lead to $m_{C} = (10^{-6} - 10^{-1})$ eV. 
 (Notice that the mixing 
parameters of active neutrinos are unconstrained by 
our scenario as they are not directly related to the mixing matrix of
singlet neutrinos.)

The above constraints imply that the condition (\ref{n1*})
is indeed satisfied in large part of the allowed parameter space. 
In particular, it holds in the two cases which we now turn to. 

As follows from Eq. (\ref{n3*}), in the case when two of the singlet
neutrinos are relatively strongly interacting the desired lepton 
(and baryon) asymmetry  $\Delta_L \sim (a~few)\cdot 10^{-9}$
is obtained for generic values of the parameters 
subject to the above constraints. For example,
for $h_A^2\sim h_B^2 = 10^{-12}$,  
$\Delta M^2 \sim M_a^2$ 
and $M_a = 10$ GeV the correct asymmetry is generated 
provided that $J \aprge 10^{-3}$ which is certainly consistent with 
Eq. (\ref{sines}). The temperature of leptogenesis is $T_L \sim 10^7$ GeV. 
In this case {\it two} active 
neutrinos are relatively heavy, $m_A \sim m_B = (a~few)$ eV, so that they 
can constitute 
the hot dark matter of the Universe. Moreover,
oscillations between them can solve either atmospheric or
solar neutrino problem, provided that their mass splitting  is small.  

A variant of our scenario makes use of
weakly interacting $N_C$ {\it and} $N_B$.  
In this case two of the usual neutrino species have
masses in the range $(10^{-6} - 10^{-1})$ eV, and the remaining one is
relatively heavy. 
As an example, let us take $h_A^2 = 5\cdot 10^{-14}$,
$h_{B}^2   = 10^{-15}$,  
$h_{C}^2 <  h_B^2$
and $M_a = 20 $  GeV, which
corresponds to the masses of usual neutrinos $m_A \sim 0.1$ eV,
$m_B \sim 2 \cdot 10^{-3}$ eV and $m_C <  m_B$. 
This variant fits particularly well into the mass
pattern suggested by the solar and atmospheric neutrino data 
\cite{Totsuka}. 
Given that $J \aprle  10^{-2}$ 
due to the constraints analogous to Eq. (\ref{sines}), 
the  correct  value of  asymmetry is obtained for 
$\Delta M^2 \aprle  10^{-2}~ {\rm GeV}^2$. 
Thus,  in this case  the singlet neutrinos should be  degenerate in mass. 
The temperature of leptogenesis is lower,
$T_L \sim 3 \cdot 10^5$ GeV. Let us note in passing
that the degeneracy of masses $M_a$ is helpful also for obtaining the
desired  baryon asymmetry 
for very small mixing angles $\theta_{ij}$.

The crucial feature of the suggested mechanism is that
it works only if 
Yukawa couplings  of all singlet neutrinos are small,
$h_a \sim (10^{-8} - 10^{-6})$.
This smallness can be explained,  {\it e.g.},
by mixing of $N_a$ with very heavy right handed neutrinos
having Yukawa couplings 
 of the same order of magnitude as those of 
quarks (and charged leptons) 
$h_q$. In this case $h_a \sim h_q \sqrt{M_a/M_R}$,
and for our values of  $h_a$ and $M_a$ the mass scale $M_R$ may be
close to the Grand Unification scale, $M_R \sim 10^{16}$ GeV.\\

We are indebted to J. Arafune and M. Fukugita for helpful discussions. 
This work was partly done during the ICTP Extended Workshop on Highlights
in Astroparticle Physics.  The work of V.R. is supported in
part by the Russian Foundation for Basic Research grant 96-02-17449a, 
and CRDF grant 649.


\begin{references}
\bibitem{*} Our analysis can be applied with
minor modifications to Dirac singlet neutrinos as well.
\bibitem{fn2} Our results for the asymmetry are valid for any other
CP-symmetric initial condition $\rho(t_i)$.
\end{references}
\end{document}